# Absence of Superconductivity in BeB$_2$

I. Felner

Racah Institute of Physics, The Hebrew University, Jerusalem, 91904, Israel

**Abstract**

The hexagonal BeB$_2$ compound has been prepared and found to be paramagnetic down to 5 K. The mixed (Mg,Be)B$_2$ system has the same T$_C$ =39 K as pure MgB$_2$, indicating that Be does not replace the Mg atoms.

israela@vms.huji.ac.il

**Introduction**

The recent discovery[1] of superconductivty around T$_C$=39 K in the simple intermetallic compound MgB$_2$ is particularly surprising for many reasons. This T$_C$ for MgB$_2$ is much higher than the highest T$_C$ values reported for the A15 intermetallic compound (Nb$_3$Ge T$_C$=23.2 K) and the borocarbides (YPd$_2$B$_2$C T$_C$=23 K). This material available from common chemical suppliers, has been known and structurally characterized since the mid 1950[2] is that of the well known AlB$_2$-type which can be viewed as an intercalated graphite structure with full occupation of interstitial sites centered in a hexagonal prism consisting of of B atoms[2]. Our scanning tunneling microscopy measurements have shown that the tunneling spectra exhibit a BCS gap structure, suggesting that MgB$_2$ is a conventional BCS (s-wave) superconductor[3].

In the search of similar intermetallic materials having high T$_C$ values, BeB$_2$ is the first natural candidate, since lighter divalent Be atoms may help providing larger phonon frequencies while keeping similar electronic properties. The crystal structure of BeB$_2$ is similar although not identical, to MgB$_2$. An earlier report[4] suggests that BeB$_2$ is hexagonal (the space group is probably P6/mmm) with the lattice parameters a=9.79(2) and c=9.55(2) A°, whereas the same space group and lattice parameters have been used[5] to define BeB$_3$. These parameters have been optimized to a =2.87 and c=2.85 A for all previous band calculations[6] of BeB$_2$, indicating that the interatomic distances should be significantly smaller than those in MgB$_2$, due to the smaller size of Be atom. On the other hand, BeB$_2$ is reported to crystallize in the AlB$_2$ type structure[7-8] similar to MgB$_2$, and other diboride compounds.

In this paper we report the magnetic properties of BeB$_2$ synthesized by arc melting. Magnetic studies show definitely that this compound is paramagnetic down to 5 K, and that superconductivity is absent. In that respect this result is consistent with the prediction of J. E. Hirsh[8], that the charge transfer from Be to B in $BeB_2$ is less than that from Mg to B in $MgB_2$, and the Fermi level in $BeB_2$ is low, beyond the regime where superconductivity occurs. It appears that in the ternary $(Mg,Be)B_2$ compounds the Be atoms do not replace Mg.

**Experimental details and results.**

Intermetalic $BeB_2$ was prepared by melting the stoichiometric elements (99.9% pure) in an arc furnace under an argon atmosphere. Precautions have been taken due to the highly toxicity of Be. Powder X-ray diffraction (XRD) measurements confirmed the crystal structure of the material (Fig. 1). The pattern was analyzed on the basis of a hexagonal structure and a least square fit of the observed peaks yields the unit cell parameters; a= 9.749(4) and c= 9.520(4) A°, in good agreement with ref. 4. The pattern contained a few unidentified extra peaks (less than 5%). The dc magnetic measurements were performed in a Quantum Design superconducting quantum interference device magnetometer (SQUID), and Fig. 2 exhibits the curve measured at 500 Oe. The curve has the typical paramagnetic shape and adheres closely to the Curie-Weiss (CW) law: $\chi = \chi_0 + C/(T-\theta)$, where $\chi_0$ is the temperature independent part of $\chi$, C is the Curie constant, and $\theta$ is the CW temperature. A fit of the CW law in the region of 5<K<150 K yields: $\chi_0 = 1.4 \times 10^{-7}$ emu/mol Oe, **$\theta = 0$ K**, and an effective moment, $P_{eff}$ equal to $0.048\mu_B$. The isothermal magnetization up to 50 kOe. is shown in the inset of Fig. 1. Based on these studies, our conclusion is that $BeB_2$ is paramagnetic.

$Mg_{1-x}Be_xB_2$ samples were also prepared by a solid state reaction as described in ref 3. Both magnetic and XRD studies indicate that Be does not enter the matrix, and the $T_C = 39$ K which is obtained is similar that of pure $MgB_2$

**Aknowledgments:**

This research was supported by the Israel Academy of Science and Technology and by the Klachky Foundation for Superconductivity.

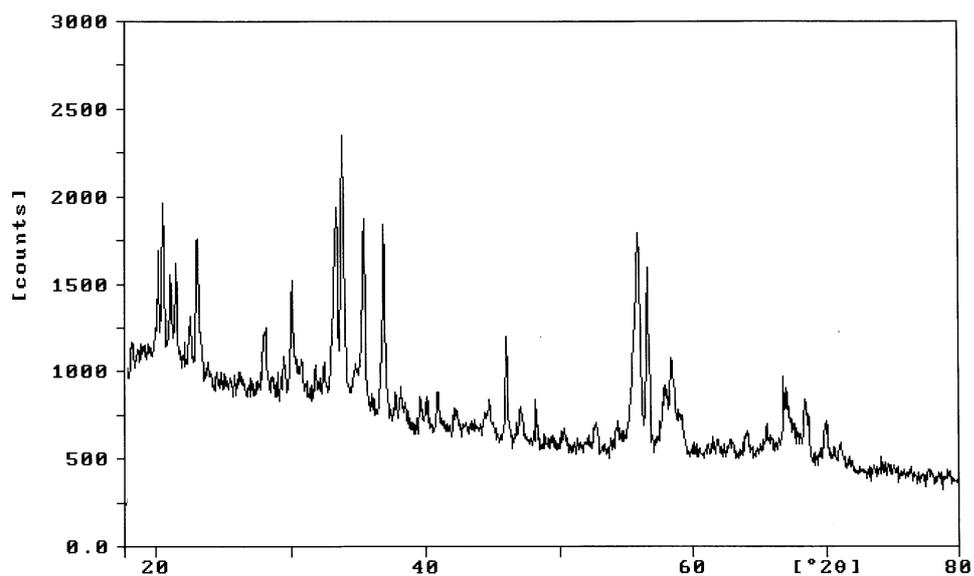

Fig.1. XRD of BeB$_2$

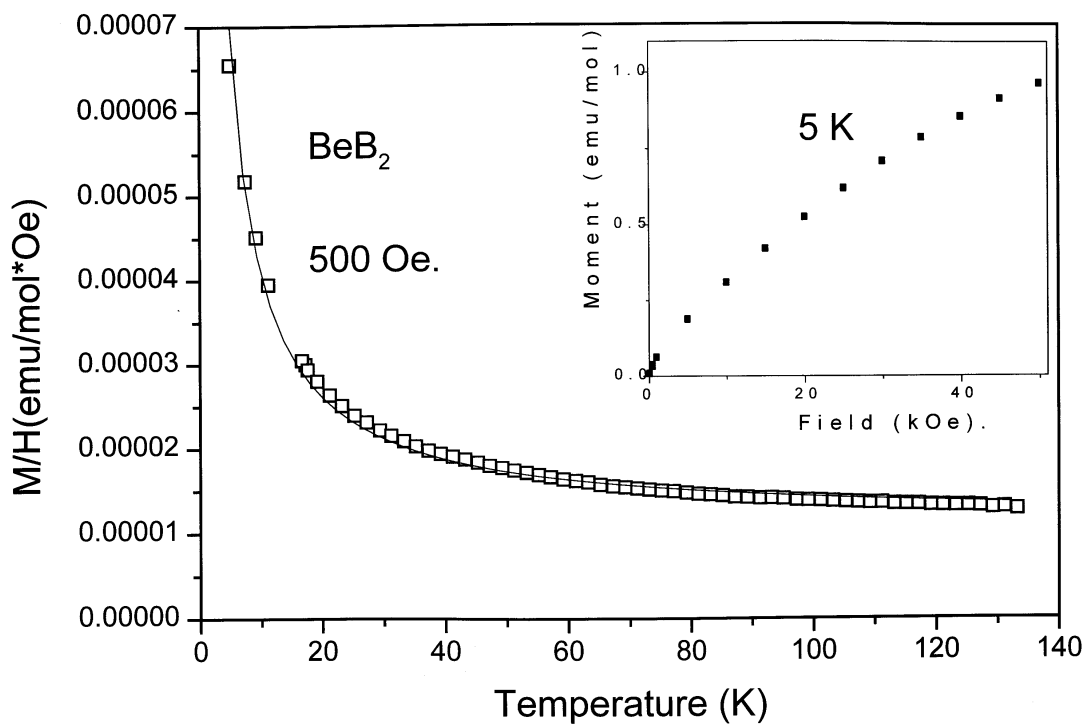

Fig. 2 Magnetic susceptibility of BeB$_2$. The isothermal M(H) curve measured at 5 K, is shown in the inset.